\documentstyle[12pt]{article}
\begin{document}
\begin{flushright}
Last Update:  09 June 1995 \\
hep-th/9502075
\end{flushright}
\begin{center}
\vfill
\large\bf{A Massive Renormalizable Abelian}\\
\large\bf{Gauge Theory in 2+1 Dimensions}
\end{center}
\vfill
\begin{center}
F.A. Dilkes\\
D.G.C. McKeon\\
Department of Applied Mathematics\\
University of Western Ontario\\
London ~CANADA\\
N6A 5B7
\end{center}
\vfill
email: TMLEAFS@APMATHS.UWO.CA\\
\hfill                     PACS No.: 11.10.Gh\\
\eject

\section*{Abstract}

The standard formulation of a massive Abelian vector field in
$2+1$ dimensions involves a Maxwell kinetic term plus a
Chern-Simons mass term; in its place we consider a Chern-Simons
kinetic term plus a Stuekelberg mass term.  In this latter model,
we still have a massive vector field, but now the interaction with
a charged spinor field is renormalizable (as opposed to super
renormalizable).  By choosing an appropriate gauge fixing term, the
Stuekelberg auxiliary scalar field decouples from the vector field.
The one-loop spinor self energy is computed using operator
regularization, a technique which respects the three dimensional
character of the antisymmetric tensor
$\epsilon_{\alpha\beta\gamma}$.  This method is used to evaluate
the vector self energy to two-loop order; it is found to vanish
showing that the beta function is zero to two-loop order.  The
canonical structure of the model is examined using the Dirac
constraint formalism.
\vfill
\eject

\section{Introduction}

It has been pointed out [1,2] that a suitable gauge invariant
action for a massive vector field in three dimensions is
$$ S = \int d^3x \Bigg[ - \frac{1}{4} (\partial_\mu A_\nu -
\partial_\nu A_\mu)^2 - \frac{\mu}{2}
\epsilon^{\mu\nu\lambda} A_\mu \partial_\nu A_\lambda + \overline
\psi (i \not\! \partial - e \not
\!\! A - m) \psi \Bigg] .\eqno(1)$$

The gauge coupling $e$ has dimension $[mass]^{1/2}$ indicating that
the theory is super renormalizable; this is borne out by the
structure of the vector propagator in a covariant gauge
$$\!\!\!\!\!\!\!\!\!\!\!\!\!\!\!\!\!\!\!\!\!\!\!\!\!\!\!\!\!\!\!\!
\!\!\!\!\!\!\!\!\!\!\!\!\!\!\!\!\!\!\!\!\!\!\!\!\!\!
D_{\mu\nu}(p) = \frac{-i}{p^2-\mu^2+i \epsilon} \Bigg[ g_{\mu\nu} -
\frac{p_\mu p_\nu}{p^2} - i\mu \, \epsilon_{\mu\nu\alpha}
\frac{p^\alpha}{p^2} \Bigg] - i a \frac{p_\mu p_\nu}{p^4}\eqno(2a)$$
$$= \frac{-i}{p^2 - \mu^2 + i\epsilon} \Big[ g_{\mu\nu} -
\frac{p_\mu p_\nu}{p^2} -
i\epsilon_{\mu\nu\alpha} \frac{p^\alpha}{\mu} \Big] - i
\Big( \frac{-i}{p^2 + i\epsilon}\Big)
\Big( - i \epsilon_{\mu\nu\alpha} \frac{p^\alpha}{\mu} \Big) - i a
\frac{p_\mu p_\nu}{p^4}\eqno(2b)$$
where $a$ is a gauge parameter.

The Chern-Simons action on its own
$$ S = \int d^3x \epsilon^{\mu\nu\lambda} A_\mu
 \partial_\nu
A_\lambda\eqno(3)$$
has itself been suggested as a suitable action for three dimensional
vector field [3]; the non-Abelian extension has been extensively
examined [4].

In this paper, we consider a model defined by the action
$$ S = \int d^3 x \Bigg\{ \frac{1}{2} \Bigg[ \epsilon^{\mu\alpha\nu}
A_\mu \partial_\alpha A_\nu + \mu \Big(A_\mu + \partial_\mu \phi
\Big)^2 \Bigg]\eqno(4)$$
\begin{eqnarray*}
+ \overline\psi \Big[\not p + e \not\!\! A - m \Big] \psi \Bigg\}
\nonumber \\
\end{eqnarray*}
\begin{eqnarray*}
\Big(\partial = ip,\, g_{\mu\nu} = (+ + -),\, \epsilon_{012} = 1,\,
\gamma_\mu \gamma_\nu = -g_{\mu\nu} + i \epsilon_{\mu \nu\lambda}
\gamma^\lambda\Big) \nonumber .
\end{eqnarray*}
The kinetic part of the action for the vector field is now the
Chern-Simons action (3); the part proportional to $\mu$  is a
Stuekelberg mass term [5].
This interaction has been considered before in [6] where it was
generated by considering a Chern-Simons plus Higgs action in the
limit where the radial component of the Higgs field decoupled.

In the next section we will discuss the quantization of this model.
The renormalization of the model will be
discussed in section three and it will be shown by explicit
calculation that to two-loop order, there is no renormalization of
either the wave function $A_{\mu}$ or the mass parameter $\mu$.
This entails using operator regularization [11], a technique which
circumvents the necessity of trying to define the tensor
$\epsilon_{\alpha\beta\gamma}$ outside of three dimensions.  The
method is first illustrated by computing the spinor two-point
function to one-loop order.  In section four the canonical structure
of (4) is analyzed.  A short discussion of the non-Abelian
generalization of (4) is in a concluding section.

\section{Quantization of the Model}

The model defined by (4) possesses the $U(1)$ gauge invariance
$$ A_\mu  \rightarrow  A_\mu + \partial_\mu \Lambda $$
$$ \!\!\!\!\!\! \phi  \rightarrow  \phi - \Lambda \eqno(5) $$
$$ \;\;\psi  \rightarrow  e^{-ie \Lambda} \psi .$$
In order to quantize this model, we add to the action the gauge
fixing term
$$ S_{gf} = \int d^3x \frac{1}{2a\mu} \Big(\partial \cdot A + a
\mu^2 \phi \Big)^2 \eqno(6)$$
where $a$ is an arbitrary gauge parameter.  With this choice, the
fields $A_\mu$ and $\phi$ decouple and the free action for the
vector field $A_\mu$ is just
$$ S_A  =  \int d^3x \, \frac{1}{2} \Bigg\{ \epsilon^{\mu\alpha\nu}
A_\mu \partial_\alpha A_\nu + \mu A^2 + \frac{1}{a\mu} (\partial
\cdot A)^2 \Bigg\}$$
$$\!\!\!\!\!\!\!\!\!\!\!\!\!\!\!\!\!\!\!\!\!\!\!\!\!\!\!\!\!\!\!\!
\!\!\!\!\!\!\!\!\!\!\!\!\!\!\!\!\!\!\!\!\!\!\!\!\!\!\!\!\!\!\!
\equiv \int d^3x \,\frac{1}{2} A_\mu M^{\mu\nu} A_\nu .\eqno(7)$$
It is easy to show that the inverse of the operator $M^{\mu\nu}$
defined in (7) is
$$ (M^{-1})_{\mu\nu} = \frac{1}{\partial^2 - \mu^2}
\epsilon_{\mu\alpha\nu} \partial^\alpha - \frac{\mu}{\partial^2 -
\mu^2} g_{\mu\nu}\eqno(8)$$
$$ \;\; + \frac{\mu(1-a)}{(\partial^2-\mu^2)(\partial^2-a\mu^2)}
\partial_\mu \partial_\nu \nonumber .$$
{}From (8) it is apparent that the propagator for the field $A_\mu$
has a pole when its momentum $p$ satisfies the equation
$p^2 + \mu^2 =0$, indicating that the vector field has a mass $\mu$.
In contrast to the propagator of (2b) there
is no long range interaction in (8). Furthermore,
the propagator behaves in leading order like $1/p$ for large
momentum, as is expected since the Chern-Simons action contains but
one derivative.  This is consistent with the model being
renormalizable since we are in three dimensions.  (We note in this
context that the gauge coupling $e$ is now dimensionless.)  If the
field $\phi$ is set equal to zero
in (4), then gauge invariance is lost and the propagator is the $a
\rightarrow \infty$ limit of (8),
which renders the theory unrenormalizable.
This $\phi \rightarrow 0$ limit of (4) and the action of (1) have
been considered in refs. [12] and [13]; indeed it was shown in [13]
that the dynamical content of these two models is the same when the
vector fields are free fields.  Once the vector fields are coupled
then the interactions of (4) and (1) are distinct in the $\phi
= 0$ gauge; the former interaction is non-renormalizable
while the latter is renormalizable.

The bilinear part of the action for the scalar field
$\phi$ has an inverse
$$M^{-1} = \frac{1}{-\partial^2 + a\mu^2} \frac{1}{\mu}
.\eqno(9)$$
This scalar field, however, is not coupled to the spinor $\psi$.
If instead of (4) we consider the interaction
$$ S_I = \int d^3x \overline\psi \Big( e_1 \not\!\! A - e_2 \not
\!\partial \phi \Big)\psi\eqno(10)$$
with the gauge transformation $\psi \rightarrow
e^{-i(e_1+e_2)\Lambda}\psi$, then $\phi$ does interact.  The
derivative coupling of $\phi$ with $\psi$ in (10) renders this
interaction unrenormalizable and hence we restrict ourselves to the
case $e_2 = 0$.

The Faddeev-Popov ghost associated with the gauge fixing of (6)
leads to a ghost propagator proportional to
$1/(-\partial^2 + a \mu^2)$;
however the ghost decouples from the remaining fields as the gauge
condition is linear in the fields and the gauge transformation (5)
is Abelian.

We now consider the renormalization of our model.

\section{Renormalization}

In order to compute radiative corrections in the model defined by
(4), we must regulate ultraviolet divergences which arise in a way
that is consistent with the three dimensional character of the
tensor $\epsilon_{\alpha\beta\gamma}$.
A variety of techniques, including
dimensional regularization, a form of Pauli-Villars and the addition
of a regulating Maxwell term to the action [2,\,7-10] have been
employed.  Perhaps the most straightforward approach is to use
operator regularization [11], a symmetry preserving procedure in
which no divergences ever appear explicitly and no regulating
parameter is inserted into the original action, thereby leaving
$\epsilon_{\alpha\beta\gamma}$ unambiguously defined.  It has been
employed in non-Abelian Chern-Simons theory [14,15,16] to one and
two loop order.

It is evident from naive power-counting arguments that ultraviolet
divergences in the theory arise in two- and three-point Green's
functions.  The photon two-point Green's function in principal can
generate divergences proportional to
$A^{\mu}\epsilon_{\mu\rho\nu}\partial^{\rho}A^{\nu}$ and
$\mu A^{\mu}A_{\mu}$; the former can be removed by renormalizing the
photon wave function while the latter actually cannot arise because
gauge invariance can be easily shown to imply that, as in four
dimensional quantum electrodynamics (QED), radiative corrections to
the two-point function must be transverse.  The spinor two-point
function is responsible for infinities proportional to
$\overline\psi/\!\!\!p\psi$ and $m \overline\psi \psi$; a spinor
wave function and mass renormalization respectively can be used to
eliminate these divergences.  The only other divergent Green's
function that can occur is the vertex function
$\overline\psi\,/\!\!\!\!A\psi$;
a renormalization of the coupling constant $e$
eliminates this infinity.  Since the form of the gauge
transformations of (5) are identical to those in QED, the same
arguments based on Ward identities [23] can be used to show that
the wave function renormalization of the photon is in fact
entirely responsible for the coupling constant renormalization.
We consequently compute the divergent
contribution to the photon two-point function.

Normally dimension regularization is the most convenient tool for
handling divergences in gauge theories.  However in the model
of (4), the intrinsically three dimensional tensor
$\epsilon_{\alpha\beta\gamma}$ occurs explicitly making it difficult
to implement this technique.  Operator regularization [11] is more
suited to this theory since no regulating parameter is ever inserted
into the initial Lagrangian leaving $\epsilon_{\alpha\beta\gamma}$
well defined at every stage of the calculation.

Background field quantization [24] is used in conjunction with
operator regularization.  The generating functional to a given order
in the loop expansion is then written in closed form, and the
logarithm of operators (at one-loop order) and the inverse of
operators (beyond one-loop order) are then regulated; the initial
Lagrangian is never altered such as by inserting a regulating
Pauli-Villars mass or by analytically continuing the number of
space-time dimensions.  We illustrate this technique by first using
it to compute the spinor self energy to one-loop order.

In order to compute this contribution to the effective action,
we first provide $\psi$ with a background piece $\chi$.
The one-loop generating functional is then given by the
superdeterminant
$$
\Gamma^{(1)}(\chi, \overline\chi) = sdet^{- \frac{1}{2}} \left(
\begin{array}{ccc} i\epsilon_{\mu\alpha\nu}p^\alpha+\mu g_{\mu\nu} +
\frac{1}{\mu} p_\mu p_\nu & -e(\gamma_\mu \chi)^T & e(\overline\chi
\gamma_\mu) \\
e(\gamma_\nu \chi) & 0 & \not\! p - m \\
-e(\overline\chi \gamma_\nu)^T & -\not \! p^T + m & 0
\end{array}\right)\eqno(11)$$
in the gauge in which $a = 1$.

We now want to extract the $(\overline\chi \chi)$ contribution to
$\Gamma^{(1)}(\chi, \overline\chi)$.  To do that, we first multiply
(11) by the constant matrix
$$
X = sdet^{- \frac{1}{2}}
\left( \begin{array}{ccc} -i\epsilon_{\mu\rho\nu}p^\rho+\mu
g_{\mu\nu} & 0 & 0 \\
0 & 0 & \not \! p^T + m \\
0 & -\not \! p - m & 0 \end{array}\right)\eqno(12)$$
so that
$$\Gamma^{(1)}(\chi, \overline\chi) = sdet^{-\frac{1}{2}} \left(
\begin{array}{ccc} (p^2+\mu^2)g_{\mu\nu}
& -e(\overline\chi \gamma_\mu)(\not \! p + m)
& -e(\gamma_\mu \chi)^T (\not \! p ^T + m)\\
e(\gamma^\kappa \chi)(-i \epsilon_{\kappa\lambda\nu} p^\lambda +
\mu g_{\kappa\nu}) & p^2+m^2 & 0 \\
-e(\overline\chi
\gamma_\kappa)^T(-i\epsilon_{\kappa\lambda\nu}p^\lambda +\mu
g_{\kappa\nu}) & 0 & p^2+m^2 \end{array}\right)$$
$$\eqno(13)$$

In operator regularization [11], we first write
$$ sdet\,H = \exp\,str\,\ln \, H\eqno(14)$$
and then regulate $\ln \, H$
$$\ln \, H = -\frac{d}{ds} \Big|_0 H^{-s}\eqno(15)$$
so that
$$\Gamma^{(1)} = \exp \frac{1}{2} \zeta^\prime(0)\eqno(16)$$
where [14]
$$\zeta(s) = \frac{1}{\Gamma(s)} str \int_0^\infty dit (it)^{s-1}
\exp - (iHt) \, .\eqno(17)$$
To extract the contribution to $\zeta(s)$ which is bilinear in
$(\overline\chi \chi)$ we employ either the Schwinger expansion
[18,11] or employ the quantum mechanical path integral [19].  Upon
identifying $H$ with the operator appearing in (13), we find that
$$ \zeta_{\overline\chi \chi} (s) = \frac{ie^2}{(4 \pi)^\frac{3}{2}}
\frac{2}{\Gamma(s)} \int d^3p \int_0^1 du\, \overline\chi(p)
\Bigg\{\frac{3 \Gamma (s-\frac{1}{2})}{[u(1-u)p^2+u \mu^2
+(1-u)m^2]^{s-\frac{1}{2}}}\eqno(18)$$
$$+ \frac{\Gamma(s+\frac{1}{2})
[\not \! p(u\mu + 2(1-u)m) + 3\mu m - 2u (1-u)p^2]}
{[u(1-u)p^2 + u \mu^2 +(1-u)m^2]^{s+\frac{1}{2}}} \Bigg \}
\chi(-p) \,.$$
As is expected, since this is a three dimensional theory, no
dependence on logarithms of $p^2$ or the masses arises in
$\zeta_{\overline\chi \chi}^\prime (0)$; such logarithms can appear
only beyond one-loop order.  Similarly, divergences appear in
renormalizable three dimensional scalar models only beyond one-loop
order [20] when using dimensional regularization.

Having seen how radiative corrections cannot result in
contributions to the renormalization group functions until at least
two-loop order when one employs operator regularization in an odd
number of dimensions, we will turn our attention to the
vacuum polarization at two-loop order.  As has been discussed above,
this will determine the two-loop contribution to the renormalization
of the coupling constant and hence will fix the beta function to
this order in perturbation theory.

The two-loop generating functional in the presence of external
vector field $A_\mu$ can be shown to be
$$\!\!\!\!\!\!\!\!\!\!\!\!\!\!\!
 \Gamma^{(2)}[A] = \frac{-ie^2}{2} \int dx dy
<x|(-i\epsilon_{\mu\rho\nu}p^\rho +
\mu g_{\mu\nu})(p^2 + \mu ^2)^{-1}|y>
\eqno(19) $$
$$\;\;\; Tr\Bigg\{\gamma^{\mu} <x|(\not\! p + e\not \!\! A)
\left[(p+eA)^2 - \frac{e}{2}
\epsilon_{\alpha\beta\lambda} F^{\alpha\beta}
\gamma ^{\lambda}\right]^{-1}|y> $$
$$\;\;\;\;\;\;\;\;\;\;\;\;\;\;
\gamma^{\nu} <y|(\not\! p + e\not \!\! A)\left[(p+eA)^2 -
\frac{e}{2}\epsilon_{\gamma\delta\sigma} F^{\gamma\delta}
\gamma^\sigma\right]^{-1}|x> \Bigg\} \,.$$
This follows from (4) and (6) upon setting $a=1$, $m=0$.

Just as (15) can be used to regulate the logarithm of an operator,
the inverse of an operator can be regulated using
$$ H^{-1} = \frac{d^n}{ds^n} \Big|_{0} \frac{s^n}{n!} H^{-s-1} \, .
\;\;\; (n = 1,2 \cdots) \eqno(20)$$
This allows us regulate $\Gamma^{(2)}[A]$ in (19) in the following
way
$$ \!\!\!\!\!\!\!\!\!\!\!\!\!\!\!\!\!\!\!\!\!\!\!\!\!\!\!\!\!\!\!\!
\Gamma^{(2)}[A] = \frac{d}{ds} \Big|_0 s
\left( \frac{-ie^2}{2} \right)
\int ^{\infty}_0 \frac{dit_1 dit_2 dit_3}{\Gamma^3 (s+1)} (it_1 it_2
it_3)^s \eqno(21)$$
$$ \;\;\;\;\; \int dx dy <x|(-i\epsilon_{\mu\rho\nu} p^{\rho} + \mu
g_{\mu\nu}) \exp -i[p^2 + \mu^2]t_1 |y> $$
$$ \;\;\;\;\;\;\;\;\;\; Tr \Bigg\{\gamma^{\mu}
<x|(\not\!p + e\not\!\!A)
\exp -i\left[(p+eA)^2 - \frac{e}{2}\epsilon_{\alpha\beta\lambda}
F^{\alpha\beta}\gamma^{\lambda}\right]t_2|y> $$
$$ \;\;\;\;\;\;\;\;\;\;\;\;\;\;\;\;\;\;\;\;\;\;
\gamma^{\nu}<y|(\not\!p + e\not\!\!A)\exp -i\left[(p+eA)^2 -
\frac{e}{2}\epsilon_{\gamma\delta\sigma}
F^{\gamma\delta}\gamma^{\sigma}\right]t_3|x> \Bigg\} \, . $$
It is possible to neglect ``regulated forms of zero" discussed in
[26-28,30] in (21) since in three dimensions one-loop subgraphs are
not divergent.  This also means that we can choose the parameter $n$
in (20) to be equal to one.

Since we are interested in only the logarithmic dependent pieces of
the two-point function, it is much easier to employ the DeWitt
expansion [25] rather than compute the full two-point function from
the Schwinger expansion [18].  The utility of this procedure at
two-loop order has previously been illustrated [16,26,27].  In this
technique we make use of the following expansion
$$ <x|e^{-i[(p+A)^2+f]t}|y> = \frac{e^{i(x-y)^2/4t}}{(4\pi it)^{D/2}}
\sum_{n=0}^{\infty} a_n (x,y) (it)^n \eqno(22)$$
in $D$ dimensions.  All dependence on $A_{\mu}$ and $f$ is contained
in the coefficients $a_n (x,y)$.  As has been argued, in order to
determine the photon wave function renormalization, the terms in the
effective action that we need to consider are bilinear in $A_{\mu}$
and contain at most one derivative of $A_{\mu}$; consequently the
coefficients $a_n (x,y)$ need to be determined only to second order
in $A_{\mu}$ and first order in $\partial_{\lambda} A_{\mu}$.  This
can be easily done using the techniques of [29]; we find that
$$ \!\!\!\!\!\!\!\!\!\!\!\!\!\!\!\!\!\!\!\!\!\!\!\!\!\!\!\!
\!\!\!\!\!\!\!\!\!\!\!\!\!\!\!\!\!\!
<x|e^{-i[(p+eA)^2-\frac{e}{2}\epsilon_{\mu\nu\lambda}
F^{\mu\nu}\gamma^{\lambda}]t}|y> \eqno(23) $$
$$  = \frac{e^{i\Delta^2/4t}}{(4\pi it)^{D/2}} \left\{
\left[1-ie\Delta\cdot A -\frac{e^2}{2}(\Delta\cdot A)^2\right]
\right. $$
$$ \;\;\;\;\;\;\;\;\;\;\;\;\;\;\;\;\;\;\;\;\;\;\;\;\; \left.
+it\left[1 - ie \Delta\cdot A\right]
\left( \frac{e}{2}\epsilon_{\mu\nu\lambda}
F^{\mu\nu}\gamma^{\lambda} \right) \right\} $$
so that to the required order
$$ \!\!\!\!\!\!\!\!\!\!\!\!\!\!\!\!\!\!\!\!\!\!\!\!\!\!\!\!\!\!\!
\!\!\!\!\!\!\!\!\!\!\!\!\!\!\!\!\!\!\!\!\!\!\!\!\!\!\!\!\!\!\!\!
<x|(\not\! p + e\not\!\! A)
e^{-i[(p+eA)^2-\frac{e}{2}\epsilon_{\mu\nu\lambda}
F^{\mu\nu}\gamma^{\lambda}]t}|y> \eqno(24) $$
$$ = \frac{e^{i\Delta^2/4t}}{(4\pi it)^{3/2}} \left\{
\frac{\not\!\!\Delta}{2t} + e \left[ -\frac{i\not\!\!\Delta
\Delta\cdot A}{2t} + \frac {i\not\!\!\Delta}{4}
\epsilon_{\mu\nu\lambda}F^{\mu\nu}\gamma^{\lambda} -
\frac{\gamma_{\mu} \Delta_{\lambda}}{2} F^{\mu\lambda} \right]
\right. $$
$$ \;\;\;\;\;\;\;\;\;\;\;\;\;\;\;\;\;\;\;\;\
\left. + e^2 \left[ -\frac{\not\!\!\Delta (\Delta\cdot A)^2}{4t}
+ \frac{\not\!\!\Delta}{4} \Delta\cdot A \epsilon_{\mu\nu\lambda}
F^{\mu\nu} \gamma^{\lambda} + \frac{i}{2}
\gamma_{\mu}\Delta_{\lambda}F^{\mu\lambda}\Delta\cdot A \right]
\right\} $$
(Here $\Delta = x-y$ and all fields are evaluated at
$z=\frac{x+y}{2}$.)

Upon substitution of (24) into (21), the two-point function, to
first order in derivatives of the external wave function, is
$$ \!\!\!\!\!\!\!\!\!\!\!\!\!\!\!\!\!\!\!\!\!\!\!\!\!\!\!\!\!\!
\!\!\!\!\!\!\!\!\!\!\!\!\!\!\!\!\!
\Gamma^{(2)}_{AA} = \frac{-ie^4}{2} \frac{d}{ds} \Big|_0 s \int
d^3 z d^3 \Delta \int^{\infty}_0 \frac {dit_1 dit_2
dit_3}{\Gamma^3(s+1)} \frac{(it_1 it_2 it_3)^{s-3/2}}
{(4 \pi i)^{9/2}} \eqno (25)$$
$$ \;\;\;\;\;\;\; e^{-i\mu^2 t_1}
e^{\frac{i\Delta^2}{4} \left( \frac{1}{t_1}
+ \frac{1}{t_2} + \frac{1}{t_3} \right) } \left
(-i\epsilon_{\mu\rho\nu} \frac{\Delta ^\rho}{2t_1} + \mu g_{\mu\nu}
\right) $$
$$ \;\;\;\;\;\; tr \Bigg\{ \gamma^\mu \left( \frac{ \not\!\!\Delta}
{2t_2} \right)
\gamma^\nu \left( \frac {\not\!\!\Delta (\Delta\cdot A)^2}{4t_3} +
\frac{\not\!\!\Delta}{4} \Delta\cdot A \epsilon_{\alpha\beta\lambda}
F^{\alpha\beta} \gamma^{\lambda} + \frac{i}{2} \gamma_\alpha
\Delta_\lambda F^{\alpha\lambda} \Delta\cdot A \right) $$
$$ \;\;\;\;\;\;\;\;\;\;\;\;\;\;\;
 + \gamma^\mu \left( - \frac{\not\!\!\Delta
(\Delta\cdot A)^2}{4t_2} +
\frac{\not\!\!\Delta}{4} \Delta\cdot A \epsilon_{\alpha\beta\lambda}
F^{\alpha\beta} \gamma^{\lambda} + \frac{i}{2} \gamma_\alpha
\Delta_\lambda F^{\alpha\lambda} \Delta\cdot A \right) \gamma^\nu
\left( - \frac{ \not\!\!\Delta}{2t_3} \right) $$
$$ \;\;\;\;\;\;\;
+ \gamma^\mu \left( - \frac{i\not\!\!\Delta(\Delta\cdot A)}{2t_2} +
\frac{i\not\!\!\Delta}{4}\epsilon_{\alpha\beta\lambda}F^{\alpha\beta}
\gamma^\lambda - \frac{\gamma_\alpha \Delta_\lambda}{2}
F^{\alpha\lambda} \right) \gamma^\nu
\left( - \frac{i\not\!\!\Delta(\Delta\cdot A)}{2t_3} \right. $$
$$ \;\;\;\;\;\;\;\;
\left. - \frac{i\not\!\!\Delta}{4}\epsilon_{\gamma\delta\sigma}
F^{\gamma\delta}
\gamma^\sigma + \frac{\gamma_\beta \Delta_\sigma}{2}
F^{\beta\sigma} \right) \Bigg\} \, .$$
In (25) we can immediately discard terms with an odd number of
factors of $\Delta_\mu$.  Remarkably, the remaining terms
proportional to $\mu^2 A^2$ and $\epsilon_{\mu\lambda\nu} A^\mu
\partial^\lambda A^\nu$ automatically cancel, eliminating the need
to evaluate any integrals explicitly or to compute any traces of
gamma matrices.  (All integrals could, in fact, be determined
using the techniques of [26,28,30].)  The fact that we obtain a
vanishing result even prior to having to compute potentially
divergent integrals indicates that the use of operator
regularization is superfluous; Pauli-Villars regularization
could also have been used to obtain this result (although it
would have been computationally more difficult).

We consequently see that no renormalization of either the vector
wave function $A_\mu$ or the mass parameter $\mu$ occurs to two-loop
order, so that the beta function and anomalous mass dimension vanish
to this order.  The vanishing of the two-loop beta function is in
accordance with the results of [31].

\section{Canonical Formalism}

We first note that the equations of motion associated with the
Lagrangian (4) are
$$\epsilon^{\mu\lambda\nu} \partial_\lambda A_\nu
 + \mu (\partial^\mu \phi + A^\mu) + j^\mu = 0\eqno(26a)$$
and
$$\partial_\mu (\partial^\mu \phi + A^\mu) = 0\eqno(26b)$$
upon varying $A^\mu$ and $\phi$ respectively. The field $A^\mu$ has
been coupled to a
classical source $j^\mu$.  If we act on (26a) with the operator
$\epsilon_{\alpha\beta\mu}\partial^\beta$, then we obtain
$$- \partial_\alpha (\partial \cdot A + \mu^2 \phi)
+ (\partial^2 - \mu^2) A_\alpha = \mu j_\alpha -
\epsilon_{\alpha\beta\gamma}\partial^\beta j^\gamma \;,\eqno(27)$$
which upon applying the gauge condition
$$\partial \cdot A + \mu^2 \phi = 0\eqno(28)$$
shows that $A_\alpha$ is indeed a field with mass $\mu$.
Furthermore, if we combine (26a) and
(26b), we see that $j^\mu$ must be conserved
(viz $\partial \cdot j = 0$).

We now show how these results can be interpreted in the context of
the canonical formalism for constrained systems as developed by
Dirac [21].  A complete formulation of the quantization of this
model has already been presented by Boyanovsky [6].
Our quantization procedure differs from that of
[6] in that we replace, in the Lagrangian,
$2A_\mu\partial^\mu\phi$ by the equivalent symmetrized expression
$A^\mu\partial_\mu\phi - \phi\partial_\mu A^\mu$.  This leads to
different expressions for $\pi_0$ and $\pi_\phi$.  Furthurmore,
in [6] the constraints associated with the momenta $\pi_i$ are
immediately classified as being second class; the first class
constraints are discussed only after the two second class
constraints are used to define the appropriate Dirac brackets
(i.e. after the corresponding variables $A_i$ are identified
as a canonical pair).
In contrast, we determine the class of the constraints in the
system by considering all four constraints simultaneously.
The physical content of the two approaches is identical.

We begin by determining the canonical momenta,
$$\!\!\!\!\!\!\!\!\!\!\!\!\!\!\!\!\!\!\!
\pi_\phi = \frac{\partial {\cal L}}{\partial(\partial^0 \phi)} =
\mu (\partial_0 \phi +\frac{1}{2} A_0)\eqno(29a)$$
$$\!\!\!\!\!\!\!\!\!\!\!\!\!\!\!\!\!\!\!\!\!\!\!\!\!\!\!\!\!\!
\!\!\!\!\!\!\!\!\!\!\!
\pi_0 = \frac{\partial {\cal L}}{\partial(\partial^0 A^0)} =
\frac{\mu}{2} \phi\nonumber\eqno(29b)$$
$$\pi_i = \frac{\partial {\cal L}}{\partial(\partial^0 A^i)} =
\frac{1}{2} \epsilon_{ij}A_j
\;\;\;\;(\epsilon_{0ij} \equiv \epsilon_{ij})\;\; , \eqno(29c)$$
from which we derive the Hamiltonian
$${\cal H} = \frac{\mu}{2} A_0^2 -\frac{1}{2\mu} (\pi_\phi -
\frac{\mu}{2} A_0)^2 + A_0
\epsilon_{ij} \partial_iA_j\eqno(30)$$
$$- \frac{\mu}{2} (\partial _i \phi + A_i)^2 - j_i A_i + j_0 A_0 .$$
It is evident that (29b) and (29c) are primary constraint equations.
By computing the Poisson bracket $\{ \pi_0 - \frac{\mu}{2} \phi,
{\cal H} \}$ we find the secondary constraint
$$\pi_\phi + \epsilon_{ij} \partial_i A_j + j_0 + \frac{\mu}{2}
A_0 = 0.\eqno(31)$$
(If the $\mu = 0$ component of (26a) is satisfied, then (29a)
and (31) are compatible.)

Unlike the corresponding constraint of ref. [6],
equation (31) cannot be identified with
the generator of gauge transformations in our approach, as it
does not commute with the constraint of (29c).
The first class constraints in our model, then, are (29b) and
a linear combination of (31) and (29c),
$$\pi_\phi + \partial_i \pi_i + \frac{1}{2} \epsilon_{ij} \partial_i
A_j + j_0 + \frac{\mu}{2} A_0 = 0 \; .\eqno(32)$$
It is easily shown that the Poisson bracket of (32) with ${\cal H}$

is zero and hence there are no tertiary constraints; no other linear
combination of constraints (29c) and (31) has this property.  The
Gauss law constraint $\partial_i E_i = 0$ in ordinary
electrodynamics is analogous to (32), as (32) generates the gauge
transformation
$$\!\!\!\!\!\!\!\!\!\!\phi \rightarrow \phi + \Lambda\eqno(33a)$$
$$A_i \rightarrow A_i - \partial_i \Lambda \; .\eqno(33b)$$
Gauge conditions compatible with (29a) and (32) are
$$\;\;\;\;\;\;\;\;\;A_0 = 0\eqno(34a)$$
$$\partial_iA_i + \mu^2 \phi = 0\;\; ;\eqno(34b)$$
these are analogous to the usual Coulomb gauge conditions in
electrodynamics.

The remaining two linear combinations of constraints in (29c) and
(31) constitute a pair of second class constraints.  We thus see
that in our model there are two second class constraints, two first
class constraints and two gauge conditions, thereby reducing the
number of degrees of freedom from eight to two: the (single)
transverse polarization of the vector and its canonical conjugate.

We note that the action of (4) is equivalent to a Freedman-Townsend
[32] type of action
$$ S = \int d^3 x \Bigg\{ \frac{\mu}{2} \epsilon^{\mu\alpha\beta}
\phi_\mu F_{\alpha\beta}(A+V) + \frac{\mu^2}{2} V_\mu V^\mu
\eqno(35)$$
$$ + \frac{m}{2} \epsilon^{\mu\alpha\beta}A_\mu F_{\alpha\beta} (A)
\Bigg\} $$
upon applying the equation of motion to the field $\phi_\mu$.  The
quantization of the action in (35) is treated in [33] using the
Batalin-Fradkin-Vilkovisky procedure [34].

\section{Discussion}

We have demonstrated that a renormalizable massive Abelian vector
theory exists in three dimensions.  Regrettably, it does not appear
possible to extend the model of (4) to
a non-Abelian gauge theory.  The replacement of the Abelian kinetic
term with a non-Abelian Chern-Simons action [4]
$$S_{CS} = \int d^3 x \, \frac{1}{2} \, \epsilon^{\mu\nu\lambda}
\Big( A_\mu^a \partial_\nu A_\lambda^a - \frac{1}{3} f^{abc} A_\mu^a
A_\nu^b A_\lambda^c \Big)\eqno(36)$$
and the Stuekelberg mass term with a Kunimasa-Goto action [22]
$$S_{KG} = \int d^3x \mu \Big( A_\mu^a - i(U^{-1}\partial_\mu U)^a
\Big)^2\eqno(37)$$
results in a gauge invariant action.  However, decoupling the field
$U$ from  $A_\mu$ through a judicious choice of gauge condition
(i.e. finding the non-Abelian generalization of (6)) does not
appear to be feasible.  Consequently it is apparently not possible
to find a renormalizable model of a massive non-Abelian gauge field
in three dimensions without invoking the Higgs mechanism.

The model of (4) is quite similar to one considered in ref. [35].
The authors of [35] have looked at the infrared limit of a vector
theory in $2+1$ dimensions defined by a Chern-Simons and Proca
mass term with a view of applying the model to anyon physics.

\section{Acknowledgments}

We would like to thank the Natural Science and Engineering Research
Council of Canada for financial support, and to Prof. A.J. Niemi for
bringing ref. [32] to our attention.

\end{document}